\documentclass{aa}

\usepackage[T1]{fontenc}
\usepackage{ae,aecompl}

\usepackage{multicol}	
\usepackage{graphicx}	
\usepackage{graphbox}
\usepackage{amsmath}	
\usepackage{amssymb}	
\usepackage{txfonts}
\usepackage{float}
\usepackage{verbatim}
\usepackage{natbib}
\usepackage{textcomp}

\newcommand\gaia{\textit{Gaia}}

\newcommand\drone{\gaia~DR1}
\newcommand\drtwo{\gaia~DR2}
\newcommand\edrthr{\gaia~EDR3}

\begin{document}
    \title{Asteroid astrometry by stellar occultations:\\ Accuracy of the existing sample from orbital fitting}
    \titlerunning{Accuracy of occultation astrometry with Gaia}


   \author{J. F. Ferreira\inst{1,3}, 
          P. Tanga\inst{1}, F. Spoto\inst{2}, P. Machado\inst{3}, D. Herald\inst{4}
          }

   \institute{Université Côte d'Azur, Observatoire de la Côte d'Azur, CNRS, Laboratoire Lagrange, Bd de l'Observatoire, CS 34229, 06304 Nice cedex 4, France \\
        \and
             Minor Planet Center -- Center for Astrophysics, Harvard \& Smithsonian, 60 Garden St., MS 15, Cambridge (MA), USA \\
        \and
             Instituto de Astrofísica e Ciências do Espaço (IA), Universidade de Lisboa, Tapada da Ajuda - Edifício Leste - 2º Piso
             1349-018 Lisboa, Portugal \\
        \and
             Trans Tasman Occultation Alliance - International Occultation Timing Association (IOTA). \\
             }

    
    \abstract
{The technique of stellar occultations, greatly enhanced by the publication of the Gaia data releases, permits not only the determination of asteroid size and shape, but also the retrieval of additional, very accurate astrometry, with a possible relevant impact on the study of dynamical properties. The use of Gaia as reference catalogue and the recent implementation of an improved error model for occultation astrometry offer the opportunity to test its global astrometric performance on the whole existing data set of observed events, dominated by minor planets belonging to the main belt.} 
{We aim to explore the performance on orbit accuracy brought by reducing occultations by stellar positions given in Gaia's Data Release 2 (DR2) and Early Data Release 3 (EDR3), exploited jointly with the new occultation error model. Our goal is to verify that the quality of DR2 and EDR3 provides a logical progression in the exploitation of occultation astrometry with respect to previous catalogues. We also want to compare the post-fit residuals to the error model.}
{We began with accurate orbit adjustment to occultation data, either alone or joined to the other available ground-based observations. We then analysed the orbit accuracy and the post-fit residuals.}
{We find that Gaia EDR3 and DR2 bring a noticeable improvement to the accuracy of occultation data, bringing an average reduction of their residuals upon fitting an orbit of about a factor of 5 when compared to other catalogues. This is particularly visible when occultations alone are used, resulting in very good orbits for a large fraction of objects. We also demonstrate that occultation astrometry can reach the performance of Gaia on small asteroids (5-8~km in the main belt). The joint use of archival data and occultations remains more challenging due to the higher uncertainties and systematic errors of other data, mainly obtained by traditional CCD imaging.}
{}
    
    \keywords{
    Occultations -- minor planets, asteroids: general -- astrometry}
    
    \maketitle
    
    \section{Introduction}

When an asteroid occults a star, its position from the observer's perspective coincides with the position of the target star. A general uncertainty exists concerning about the exact location of the star with respect to the asteroid's centre of mass during the interval of time in which it is hidden. This uncertainty, related to the asteroid shape and size, must be carefully evaluated, but of course its amplitude in absolute terms is reduced in proportion to the size of the occulting body.

As the \gaia{} mission by ESA improves star positions and asteroid orbits, reliable predictions become available to exploit a large amount of possible events, even from a single location. We recently studied the performances of an astrometric survey systematically run by a single robotic telescope \citep{Ferreira_2020} that collects only isolated occultation "chords" (i.e. a single "cut" along a segment crossing the projected asteroid shape on the sky). Coordinate observations by multiple stations, observing the occultation produced by different chords of the same object, are also commonly obtained, thanks to a wide network of observers involving both amateur and professional astronomers \citep{Herald_2020}. In this case, of course, the uncertainty on the location of the asteroid relative to the star is reduced, compared to single-chord events. Such multi-chord events are already routinely exploited to optimise predictions for events of specific targets, and to derive physical data on TNOs \citep{varda}, Centaurs \citep{RECON}, and Jupiter Trojans \citep{leucus}. Each of the referenced results are examples of a larger set of papers focused on each of these groups.

As a large majority of the occulted stars are today in the \gaia{} data releases (exceptions can exist due to the marginal incompleteness of the catalogue, especially for bright stars), the resulting asteroid astrometry (hereinafter occultation astrometry) is very accurate. In \citet{spoto_17}, we showed the impact of \drone{} \citep{gaiaDR1} on the occultation astrometry, finding that, even with this preliminary accuracy, some orbits derived from data of occultations only on the Minor Planet Center (MPC) website\footnote{\url{https://minorplanetcenter.net/}} can already be of better quality with respect to orbits obtained from all available data (mainly ground-based CCD imaging). The gain in accuracy can reach one order of magnitude or more, depending on how many occultation events are available for a given object. It should be noted that this result was just a proof of concept on the capability of Gaia astrometry to make occultations self-consistent in terms of orbital precision, but it was still very preliminary both in terms of the stellar accuracy provided the mission, and the extent of the analysis of the results.

\drtwo{} \citep{gaiaDR2}, represents a major step with respect to \drone{} and the first to reach the domain of accuracy and completeness expected from \gaia{} with the inclusion of parallaxes and proper motions as a major step forward in the exploitation of occultation astrometry. \edrthr{} \citep{gaiaEDR3} goes further in terms of data cleanness, precision and absence of systematic errors. With respect to \drone{}, a full order of magnitude improvement on star positions has been realised. 

This advantage has now been exploited in several situations, but two extreme examples can be considered emblematic: the TNO Arrokoth \citep{arrokoth}, the second target of the NASA New Horizons mission, for which the occultation was fundamental to the successful fly-by, and, at the opposite end of the distance scale, (3 200) Phaethon, which is the target of the JAXA/DESTINY+ mission, with a whole set of occultations successfully predicted and observed over the year 2019 \footnote{Example:\url{http://www.euraster.net/results/2019/index.html\#1015-3200}}. These cases, which were beyond any possibility just a few years ago, illustrate the major breakthrough brought by the new era of astrometry well. They allow predictions for bodies that are either extremely small in apparent size and/or have high apparent motion. Also, they further illustrate how detecting a first event of a difficult target and obtaining its astrometry is beneficial to the improvement of the predictions for subsequent events. 

While these new releases have effectively boosted the accuracy of occultation predictions, here we focus on the strong impact of this improved astrometry obtained from past events. Several articles confirm the effectiveness of occultation astrometry (for a recent example see \citep{Rommmel}) but none of them propose a global statistical analysis on all the existing record, in particular for the large data volume represented by the main belt. With this work, we intend to fill this gap. 

In the process, we also investigated the statistics regarding orbit accuracy when occultation astrometry is used jointly to other astrometric measurements. This is an essential step to obtain the best possible orbits to study the NEA impact risk and secular orbit changes induced by the Yarkovsky effect. In this context, we want to clarify the role of occultation astrometry with respect to asteroid astrometry obtained by Gaia itself, as occultations may represent the only technique capable of extending the ultra-accurate astrometry of asteroids collected by Gaia itself, by collecting data of equivalent quality beyond the duration of the mission. This possibility further stresses the importance of having a clear view of the statistical properties of occultation astrometry.

Another point of interest is that the use of the stellar astrometry by Gaia is sometimes not straightforward as several quality indicators and error sources appear in the tables, which can be relevant in specific cases. We assessed the role of the most relevant indicators in the frame of occultation astrometry.

Our article is organised as follows. We first describe (Sect.~\ref{S:errmodel}) the general properties of the data set of the available occultations, from which the properties of the associated error model derive. In Section \ref{S:errstar} we discuss the role of the uncertainties related to stellar astrometry and the use of quality indicators in the Gaia data releases.
We then determine an orbital fit to the occultation data and extract their residuals (Sect.~\ref{S:residuals}). We conclude by discussing the quality of the astrometry with respect to Gaia and the challenge of combining asteroid occultation data with the large amount of astrometry available from the ground (Sect.~\ref{S:discuss}). 

\section{Data properties and error model for the occultations}
\label{S:errmodel}

Main belt asteroids are the most represented population in the data set of the currently available occultation astrometry \citep{Herald_2020}, so they constitute a relevant statistical sample to assess its performance. 

Observations of stellar occultations are regularly updated and distributed through the Planetary Data System \citep{herald_2019} and (with more frequent updates) through the database maintained by the same author and associated with the \texttt{Occult} package\footnote{\url{http://www.lunar-occultations.com/iota/occult4.htm}}. They are also submitted to the MPC. However, currently, not all the available information is accessible through the MPC (most notably, the error model). We thus made use of the full astrometric output available directly from the data reduction. 

The whole past record of occultation astrometry was recently reprocessed by \citet{Herald_2020}  to take into account \gaia{} accuracy effects (such as relativistic light deflection, for the star and the asteroid separately) and to adopt a more sophisticated error model capable of taking into account the different situations that are encountered in the data. We invite the interested reader to refer to that reference for a complete review of this complex process, including the definition of the error model.

\begin{figure*}[t]
\includegraphics[width=1.\hsize]{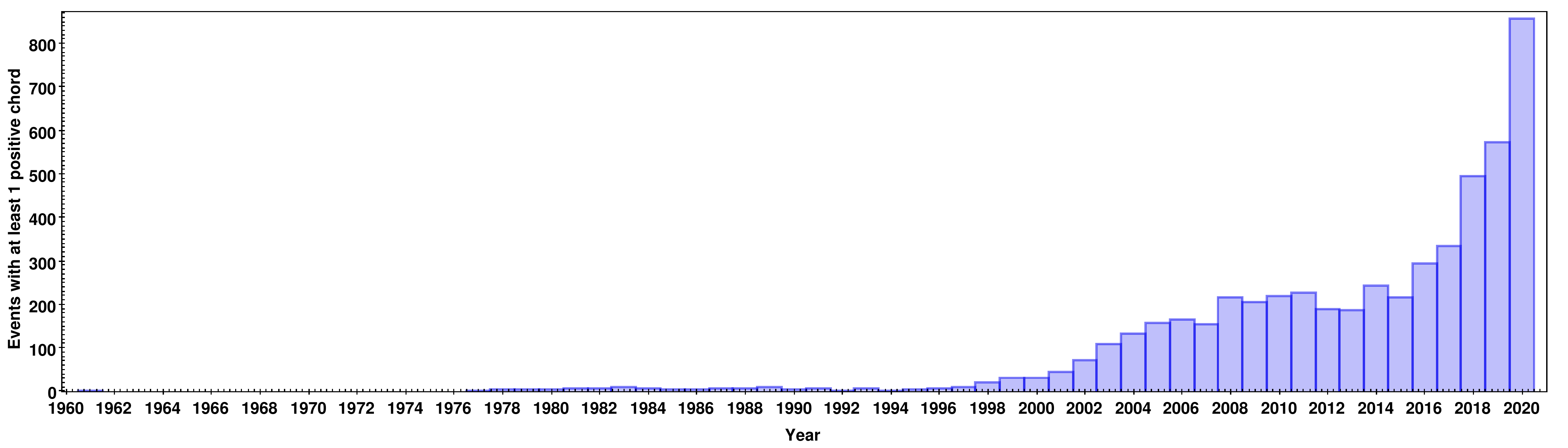}
\caption{Number of positive occultation events in the data set of asteroid occultations, per year. Earlier isolated observations before 1980 are not represented in this plot. We remark here the sharp increase in the number of events starting in 2018 is a consequence of better predictions provided by \drtwo{} that year. At this point, it is too early to notice if there is a similar trend thanks to EDR3, as we have only recently begun to receive data from those updated predictions.}
\label{F:occperyr}
\end{figure*}

In Fig.~\ref{F:occperyr}, we show the distribution of the number of observed occultations ('positive' events mean that at least one chord was reported) as a function of time. The first striking feature is the sharp increase during the 1990s, due to the release of the astrometry by the Hipparcos mission \citep{hipparcos}, alongside to the diffusion of affordable electronic cameras and GPS devices. A relative stagnation between 2005 and 2015 follows, during which asteroid orbits continue to improve but very gradually, due to accumulating astrometry. At the same time a slow deterioration of star positions due to proper motion uncertainties in Hipparcos starts to show up. Then, a new season begins in 2016 (\drone{}), followed by an acceleration in 2018 (\drtwo{}).

All along this time frame, the improvement in predictions is also accompanied by important changes: in the acquisition techniques (from visual timing to video cameras), time tagging (from manual chronometers, synchronised to audible reference signals, to GPS), and the number of active observers, just to cite the relevant ones. Given the relatively recent surge in the number of observations, it is clear that overall statistics will be dominated by this last period, characterised by evolved equipment and techniques. 

\begin{figure*}[t!]
\includegraphics[width=1.0\hsize]{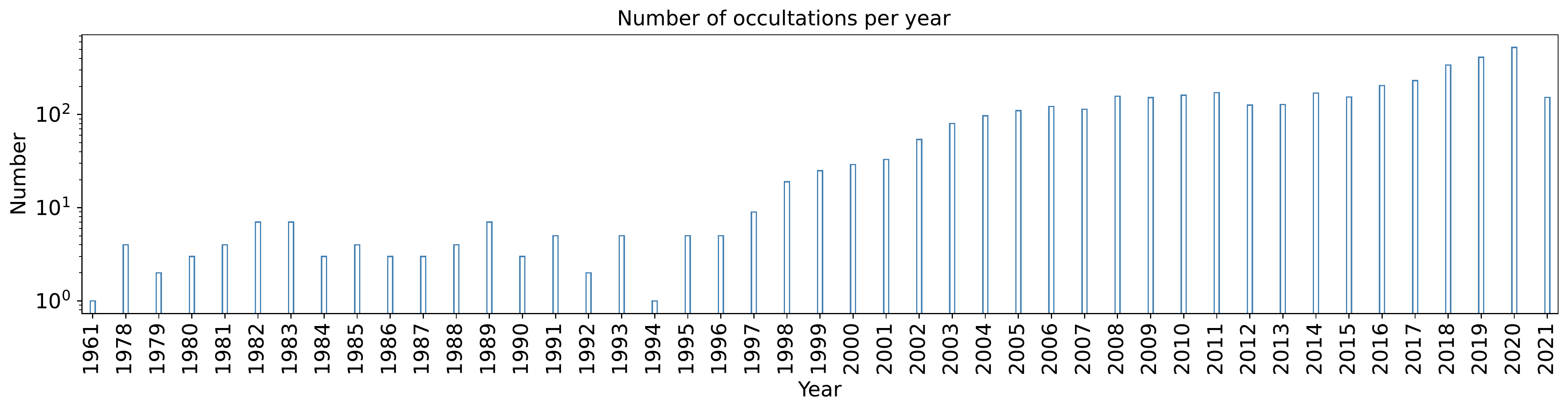}
\includegraphics[width=1.0\hsize]{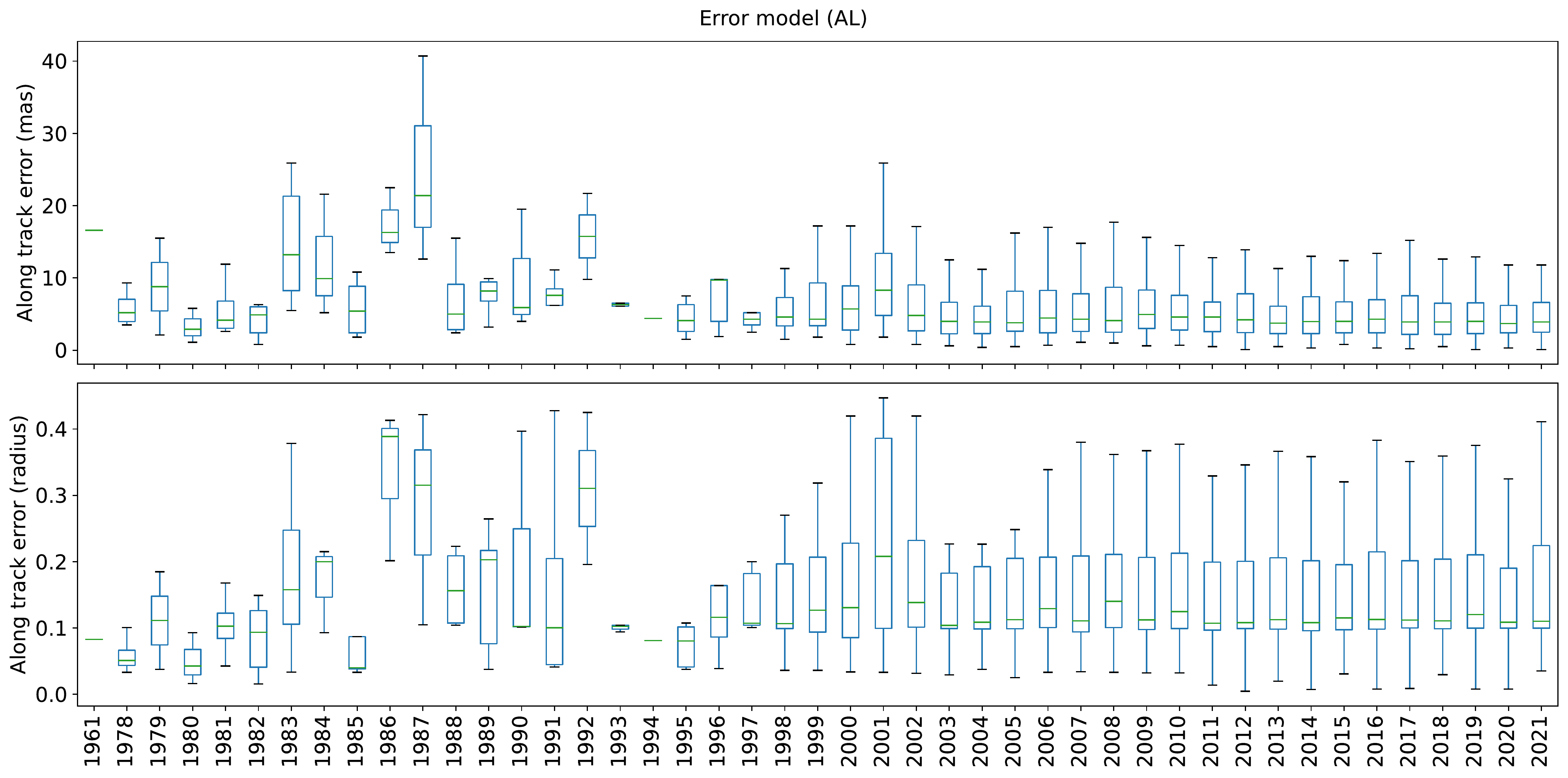}
\includegraphics[width=1.0\hsize]{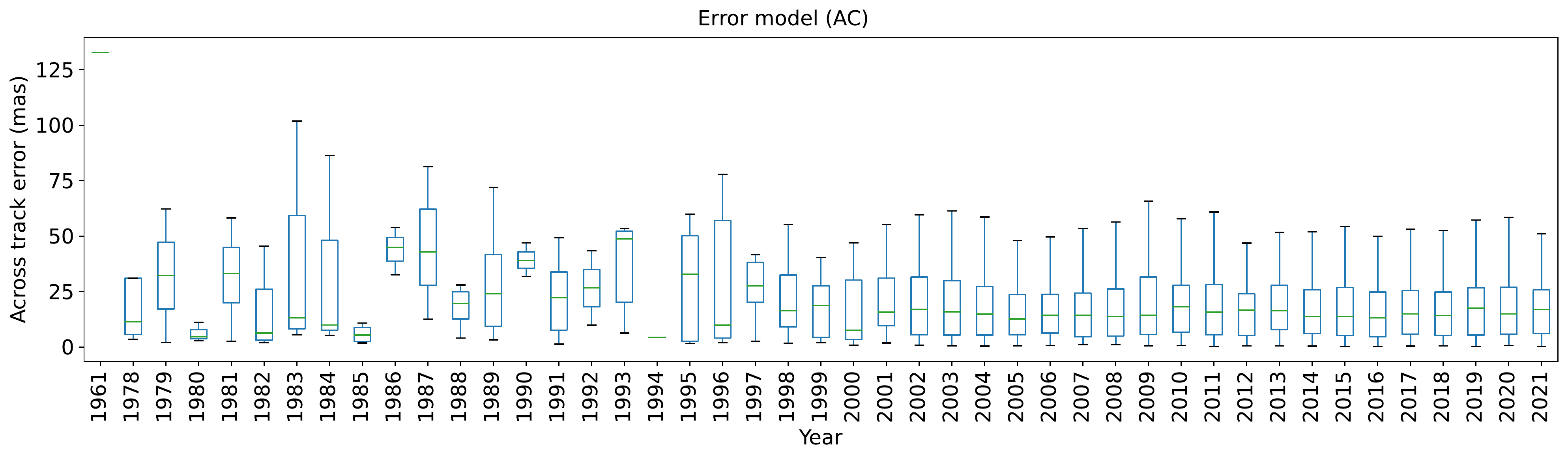}
\caption{Statistics of the along-track (AL) error plotted against the year of the observations. It incorporates several effects, including timing errors, uncertainties due to size and shape models (particularly relevant for single-chord events, which represent 61$\%$ of the total). Relevant features are the high dispersion before the late 90s, and the overall consistency afterwards. The AL error (two middle panels) is provided both in absolute values (mas) and in units of the asteroid radius. For reference the first row repeats the plot with the number of occultations for each year.}
\label{F:AL_error_model}
\end{figure*}

\begin{figure*}[t!]
\includegraphics[align=c,width=\hsize]{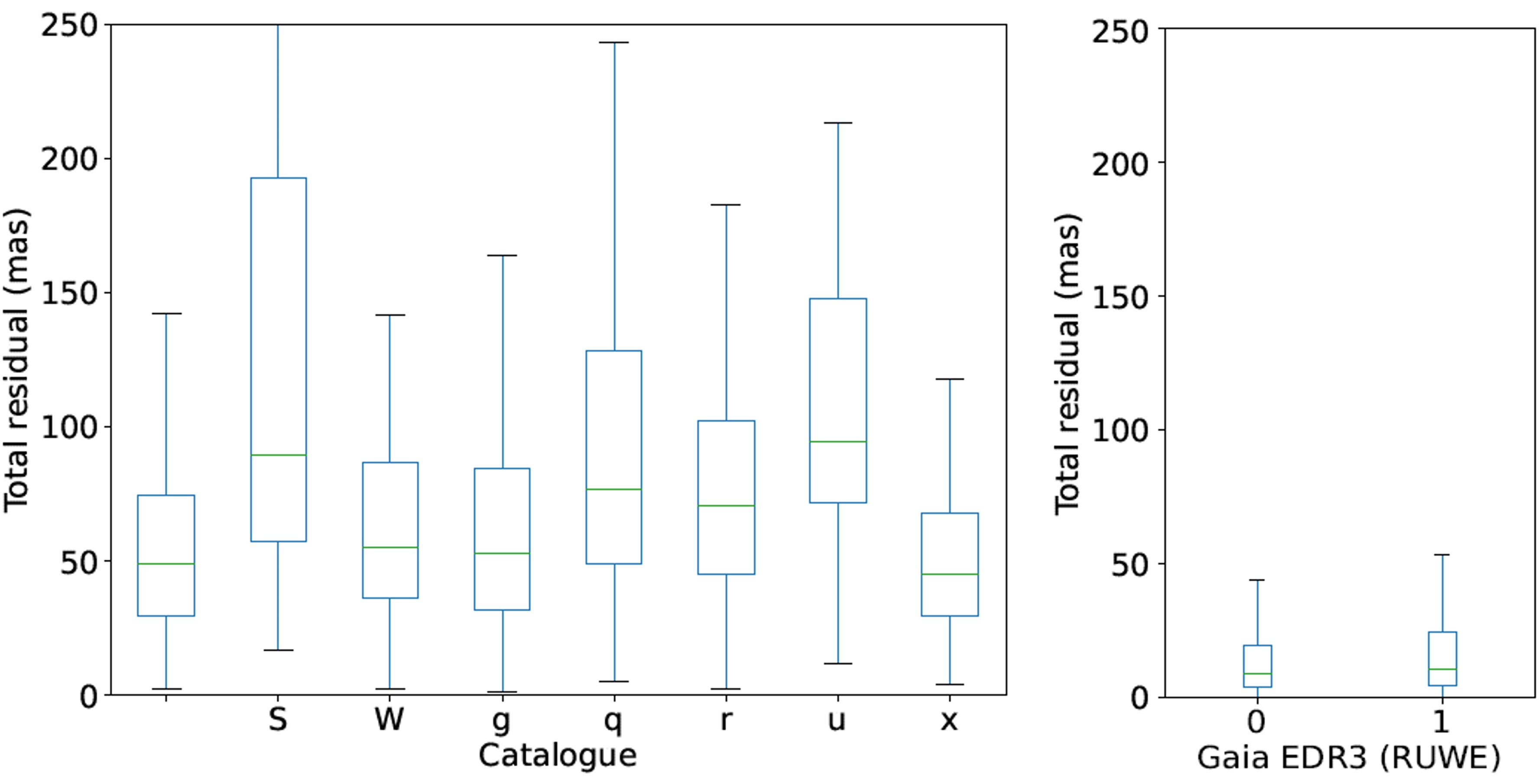}
\caption{Residuals on occultation astrometry (in mas), reduced with respect to pre-Gaia catalogues, obtained by fitting all observations at MPC (left panel). MPC catalogue codes are the following:  s: USNO B2.0; w: CMC-14; g: Tycho-2; q: UCAC-4; r: UCAC-2; u: UCAC-3; x: Hipparcos 2. No letter indicates an unknown catalogue. Right panel: Occultations reduced with \edrthr{}. The values 0 and 1 on the horizontal axis correspond to \texttt{RUWE}$\,<1.4$ and \texttt{RUWE}$\,\geq 1.4$, respectively.}
\label{F:residual_catalogues}
\end{figure*}

This evidence, however, does not hide a certain heterogeneity even in recent data, as observational problems (due to incorrect calibration, non-standard equipment behaviour, difficult meteorological conditions, incorrect identification of the target star, etc.) or technical differences (for instance, time tagging by a network time protocol or one-pulse-per-second GPS signal) continue to be present and represent a source of dispersion in the uncertainty of the observations.

In addition to such effects, uncertainties on occultation astrometry come from limitations that are intrinsic to this technique. Most notably, the derivation for an event of the position of the centre of mass of the occulting body (referred to a conventional epoch for the event) from the observed occultations chords, is a process affected by uncertainties related to the number and accuracy of the available chords, the knowledge on the asteroid shape and size, and the position of the centre of mass in the projected asteroid silhouette. For obvious reasons, the amplitude of these errors increases proportionally to the object size. 

The final source of uncertainty is the stellar astrometry available from the catalogues, which is propagated to the epoch of the event and treated separately in the next section.

For all purposes of orbit determination, the use of an appropriate error model, translating into weights on each astrometric point, is fundamental to optimise the results \citep{carpino}. 

The recently revised heuristic error model provided by \citet{Herald_2020} is based on an estimation of the uncertainty sources. It also incorporates lack of knowledge on the object shape, size, and mass distribution. As we adopted it for the orbital fit in this article, it can then be compared to the residuals with respect to the best fitting orbit. Of course, as the given error estimates are empirical, we cannot expect them to be reproduced by the post-fit residuals exactly. However, this comparison can potentially lead to future improvements of the error model, which are beyond the goal of this work.

Hereinafter, we address the uncertainties and residuals of occultations by projecting them onto two orthogonal directions, defined on the sky plane, for each event. These directions are across-track (AC) and along-track (AL), where the track direction is defined by the instantaneous proper motion of the asteroid as seen by the observer of the occultation. The AL direction is parallel to the object's movement, and it is mostly associated with timing uncertainties. The AC direction, on the other hand, is perpendicular to the movement, and it is associated with the uncertainty of the path's position on Earth. This is the same scheme illustrated in \citet{Ferreira_2020}.

One should note that depending on how the occultation astrometry is reduced, it can be referred to a topocentric location of an observer, or a conventional point in space (for instance, the geocentre or another topocentric position). For the sake of our analysis, as long as the position and the proper motions are consistently computed for the chosen reference frame, our results will not change.


Figure \ref{F:AL_error_model} shows the distribution of the errors derived by the model above, over time, separating the contribution in the AC and AL direction. Of course, AL presents the highest accuracy. A plot is also provided with the uncertainty normalised to the asteroid radius as determined by the WISE survey \citep{mainzer_2011a, mainzer_2011b, masiero_2011}. This version allows us to appreciate the uncertainty on the position relative to the object size. As mentioned above, some older occultations can be less reliable, and single chords are, by nature, less reliable than multi-chords.

The small number of observations and chords per event is responsible for the large fluctuations till the mid 1990s. Afterwards, the effect of a more homogeneous, standardised approach and increasing statistics appear, with a remarkable stabilisation of the performances. The average of the distribution fluctuates around 3-4 mas AL. It is interesting to note that in this last part of the time frame, the AC values have a larger spread, but they still achieve a very interesting performance of the order of $\sim$10 mas. 

\section{Errors on stellar astrometry}
\label{S:errstar}

While the error sources represented above are intrinsic to the occultation technique, absolute astrometry is also affected by the additional uncertainty on the position of the occulted star. This of course depends strictly on the source of information, the advantage being that new astrometric catalogues bring progressive improvements to our knowledge of positions and allow us to correspondingly refine occultation astrometry. For practical use with occultations, we are concerned here by the astrometric error propagated to the epoch of each event. 

The catalogue position of the target star and their uncertainties are propagated by using the provided information on proper motion and parallax given by each catalogue. In the case of \edrthr{} the rigorous epoch propagation is applied, as in Appendix C of \citet{but_lin_2014} and in the \edrthr{} online documentation\footnote{\url{https://gea.esac.esa.int/archive/documentation/GEDR3/Data_processing/chap_cu3ast/sec_cu3ast_intro/ssec_cu3ast_intro_tansforms.html}}. The error on the star position is then summed quadratically to the error of the occultation. This procedure is also explained in \citet{Herald_2020}.

The orbital fit is obtained by the OrbFit\footnote{\url{http://adams.dm.unipi.it/orbfit/}} software tool and optimised in order to take into account the complete error model for occultations. In the process, the normalised $\chi^2$ value of each observation is used to reject outliers, defined by $\chi^2 > 10$.

To show the jump in precision brought by the recent Gaia data release, we compare the performance of \edrthr{} on occultation astrometry, to the pre-Gaia situation when several catalogues were used to reduce both occultation results and ground-based measurements.

The striking difference is illustrated by Fig.~\ref{F:residual_catalogues}, showing the distribution of the residuals to the best fitting orbit for all objects.

The left panel shows the statistics on post-orbital fit residuals obtained for occultations. The fit considers all optical astrometry available, as provided by the MPC database, for the 1616 asteroids that have occultation data. The total number of occultations used is 5773. 
For archive data, OrbFit uses a new error model that is also being tested under this updated version, which estimates the uncertainty of observations based primarily on year of observation, observatory code, star catalogue used and asteroid magnitude. This new error model has not yet been published, but an article is in preparation by Spoto et al. and is expected to be published in 2022.

As shown by the plot (left panel), the typical average accuracy with recent pre-Gaia catalogues is around 30-50 mas, and it reaches 80-90 mas for some events associated with specific astrometric catalogues.  In the right panel, \edrthr{} has been used for the star positions. The gain in accuracy, reaching approximately the 10 mas level, is clearly visible. To investigate the role of quality indicators in \edrthr{} we considered the two main quantities that can suggest potential issues in the astrometric solution, namely the renormalised unit weight error (RUWE)\footnote{\url{https://gea.esac.esa.int/archive/documentation/GDR2/Gaia_archive/chap_datamodel/sec_dm_main_tables/ssec_dm_ruwe.html}} and the \texttt{duplicated\_source flag}\footnote{\url{https://gea.esac.esa.int/archive/documentation/GDR2/Catalogue_consolidation/chap_cu9val_cu9val/sec_cu9val_942/ssec_cu9val_942_dupl.html}} flag. 

The \texttt{RUWE} is the reduced $\chi^2$ value of the astrometric solution, further normalised by a function of the star brightness and colour index, determined from the statistical properties of the astrometric errors. The related study \citep{Lindegren18} defines different levels of \texttt{RUWE} thresholds as a function of the degree of cleanness desired in a sample of stars. The conventional threshold that appears to be optimal is \texttt{RUWE}=1.4, adopted here for the plot in Figure~\ref{F:residual_catalogues} (right panel). \texttt{RUWE} turns out to be more reliable in general than other quality indicators taken separately, proposed in the Gaia archive.

\texttt{duplicated\_source} is a flag, set when the star could potentially have a secondary source deserving a special data reduction (that has not been applied in the currently available releases. Potentially (but not systematically, this can lead to biases in the astrometric solution, of unknown amounts).

In principle, RUWE and \texttt{duplicated\_source} are derived independently during the astrometric reduction of Gaia data, but the potential duplicity indicated by \texttt{duplicated\_source} for a source can bring an increase to its RUWE value. A random sampling of Gaia EDR3 with three million stars of magnitudes below 16 (the usual occultation targets) showed that 25\% of stars with a duplicate source flag have a RUWE above 1.4 and this value drops to 19\% for stars without a duplicate source flag. So, the correlation  exists, but it remains rather weak and can be neglected in this context.

All the occulted stars that we considered have non-zero \texttt{RUWE} values, implying that the astrometric solution provides at least five parameters (position, parallax, proper motion). Both \texttt{RUWE} and \texttt{duplicated\_source} can indicate the possible presence of unknown systematic errors that can affect the astrometry of a source, which is larger than its nominal uncertainty. We find that \texttt{RUWE} is indeed significant as shown, with a small difference in the statistics of the residuals of the occultation astrometry. We do not show a similar comparison here with respect to \texttt{duplicated\_source} as no difference can be detected, which is a possible sign that this might be a secondary source of errors, which appears for a very minor fraction of the target stars. 

\section{Orbital fit and residuals}
\label{S:residuals}

We then exploited the residuals computed by the orbital fit of occultations alone, reduced with the stellar astrometry by \edrthr{}. For this task, we selected occultations of stars with \texttt{RUWE}$\,<1.4$, starting from the year 2000. By using this filter, we eliminated a very small fraction of the sample where, as mentioned above, observations are less reliable and homogeneous. We then selected objects that have at least four occultation events, as a minimum to grant a meaningful comparison with asteroids that have a larger number of occultation events. This left us with 512 objects available for analysis, with the number of occultation events for a single object reaching a maximum of 26.

Of course, a very small number of astrometric data points, when just a few occultations are available, does not mean that the best possible orbit is obtained. The formal uncertainty of the fit can be small, but some systematic differences can be present with respect to a fit to a large astrometric data set. However, for the goal of showing the self-consistent accuracy of occultation data, this is not a limitation.

\begin{figure}[ht]
\includegraphics[width=0.95\hsize]{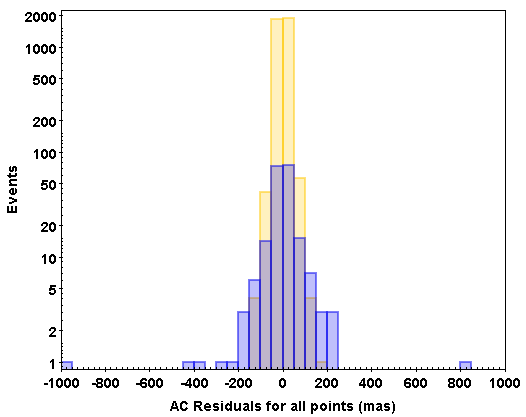}
\includegraphics[width=0.95\hsize]{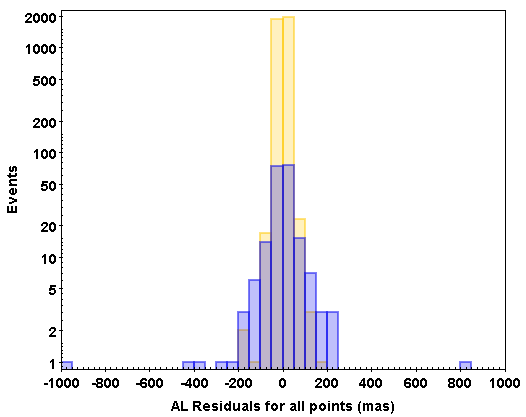}
\caption{Distribution AC and AL residuals for the astrometry obtained by occultations, when orbital fitting is run on them alone. The dark (blue) colour and the adopted log scale emphasise outliers that are rejected by the fit procedure. }
\label{F:hist_residual}
\end{figure}

For a more significant representation of the results, residuals are projected on the (AL, AC) plane. 
The histograms representing the distribution of residuals in the AC and AL directions are given in Fig.~\ref{F:hist_residual}. The core of the distribution is clearly concentrated in the central peak, whose full width at half maximum is around 10 mas for both AL and AC.  

One should also note that the queue of the distributions contain a small number of rejected observations. The fraction of rejection is rather important for years $<$2000 (20.0$\%$), but decreases afterwards (5.5$\%$). As this last time interval contains $\sim$25 times more events than the former, it largely dominates the statistics.

\begin{figure*}[th]
\includegraphics[width=0.45\hsize]{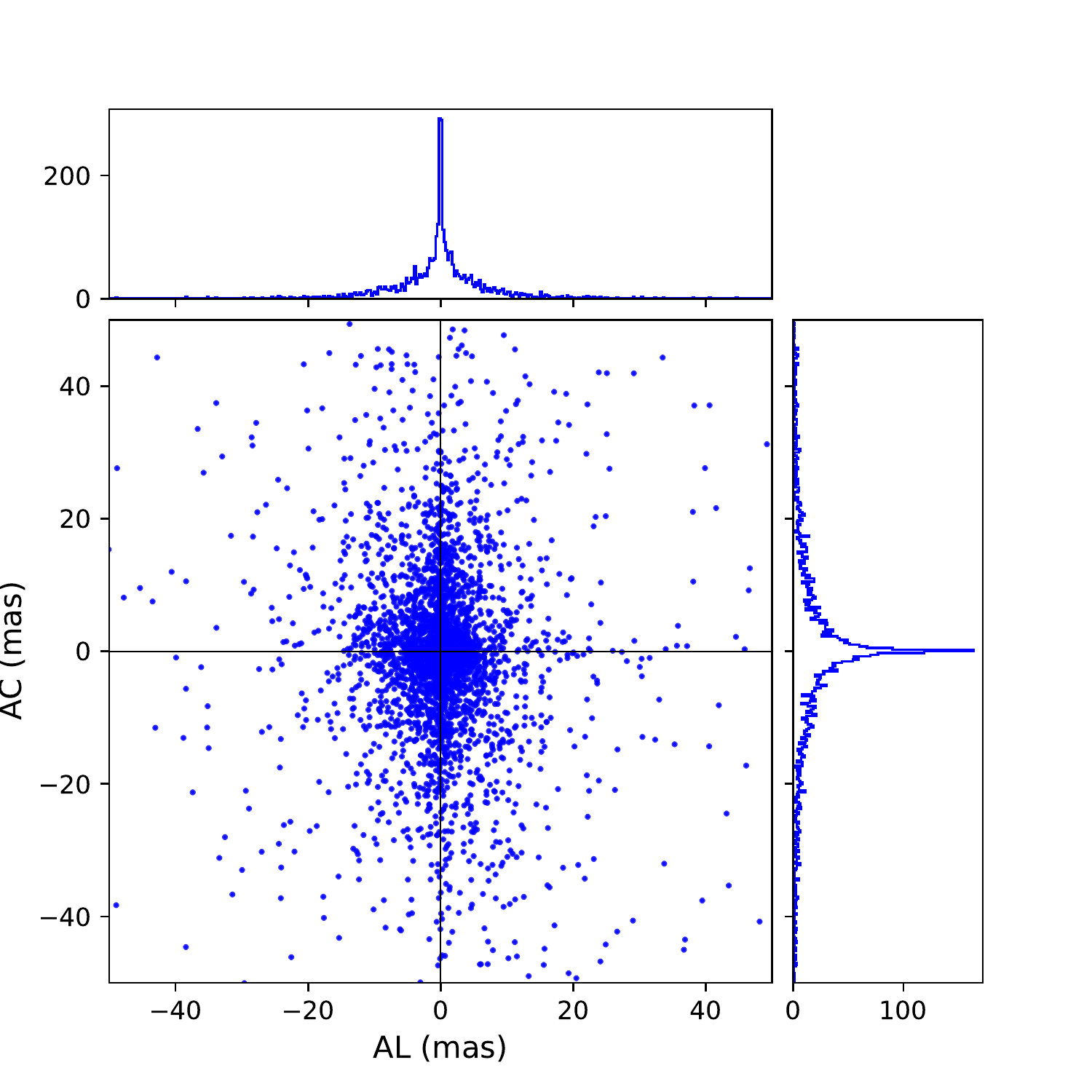}
\includegraphics[width=0.45\hsize]{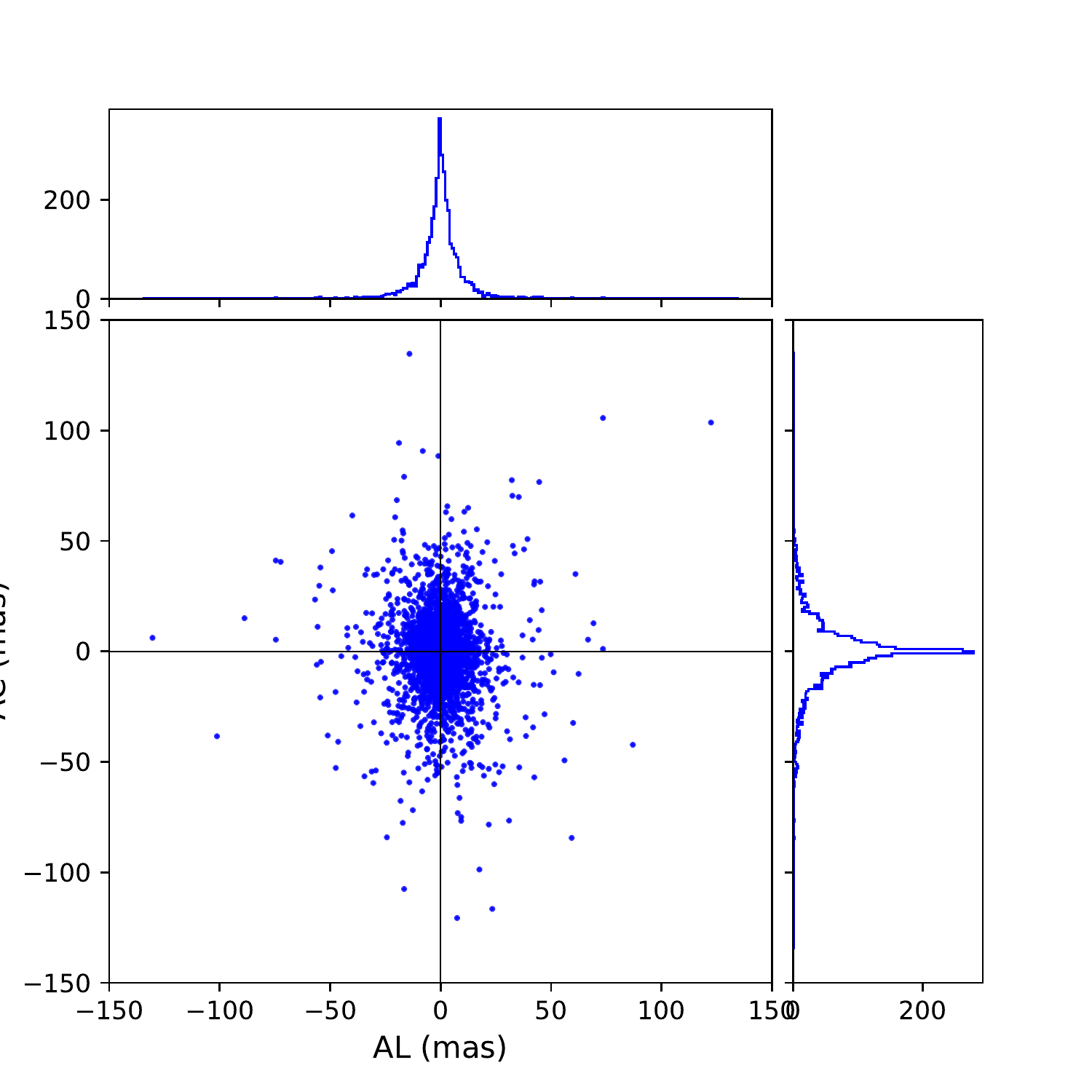}
\caption{All residuals for orbits fitted to occultations only (left panel) and to all the optical astrometry available from the MPC (plus occultations). The occultation data considered here are in years $\ge$2000. Bin size for the histograms is 1 mas.}
\label{F:post2000}
\end{figure*}

\begin{figure}[ht]
\includegraphics[width=1.\hsize]{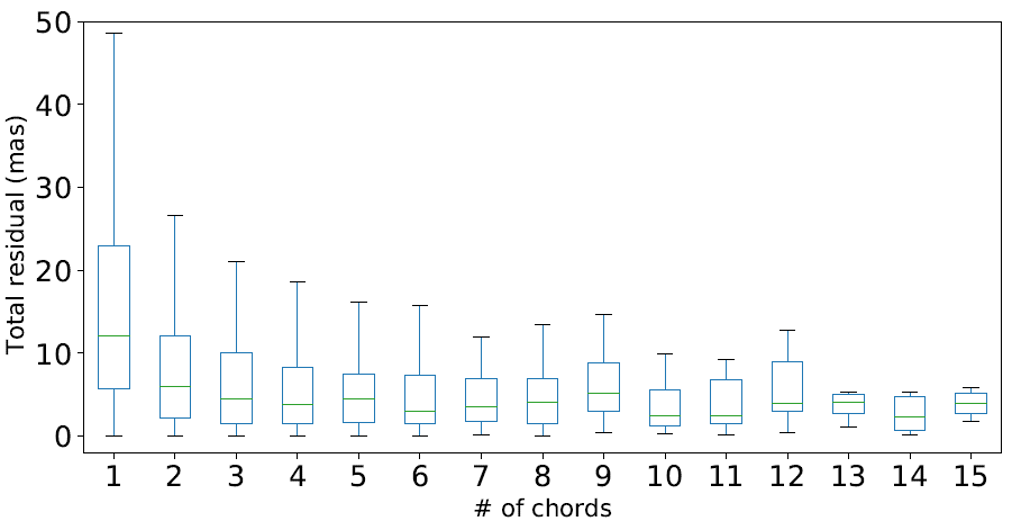}
\includegraphics[width=1.\hsize]{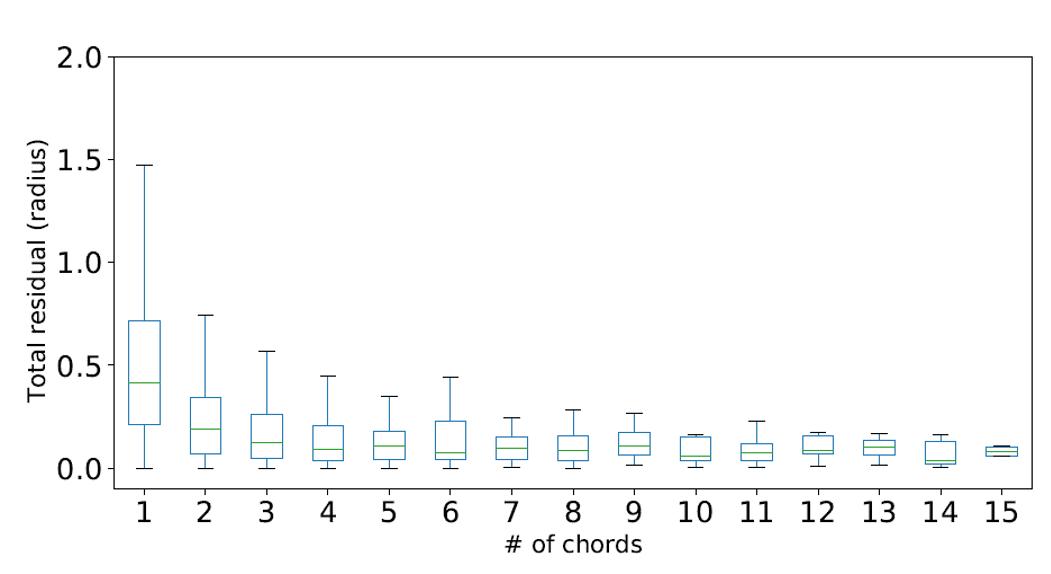}
\caption{Distribution of total residuals on occultation astrometry (quadratic sum of AC and AL residuals), obtained by fitting the astrometry derived by occultations only, as a function of the number of chords. All occultations have been reduced by adopting stellar positions available in  \edrthr{}. A few events with stars not in \edrthr{} have not been considered. Top panel: Residual in mas. Bottom: Residual normalised to the apparent asteroid radius at the epoch of the event.}
\label{F:AL_residual_catalogues}
\end{figure}

Another representation of the residuals is provided in Fig.~\ref{F:post2000} and is limited to non-rejected astrometry obtained after the year 2000, where data are more homogeneous. As expected, there is a trend in the statistics of the residuals of occultations as a function of the number of observed chords (Fig.~\ref{F:AL_residual_catalogues}), but it is rather subtle and - beyond 6-7 chords, partially masked by fluctuations due to the drastically decreasing number of events. 

\begin{figure}[ht]
\includegraphics[width=0.95\hsize]{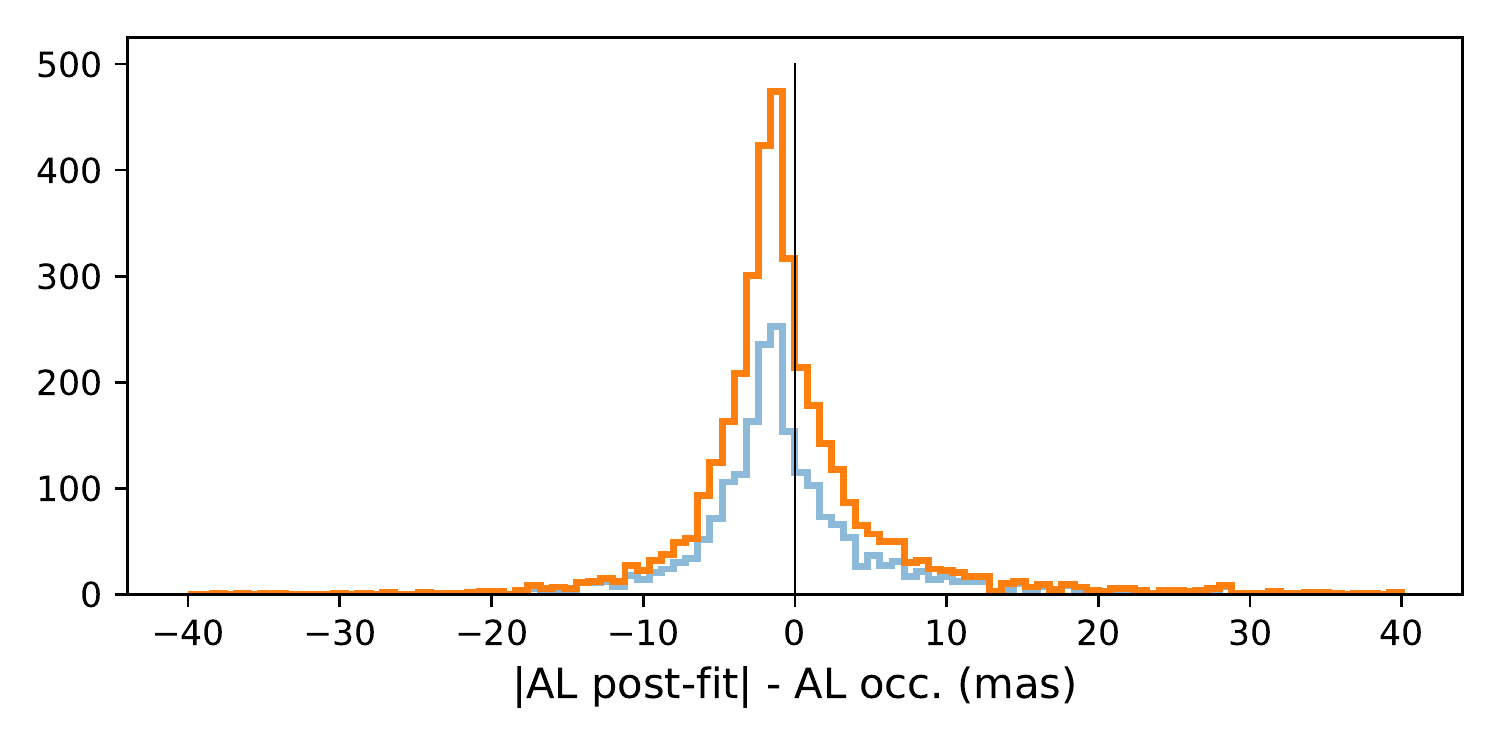}
\includegraphics[width=0.95\hsize]{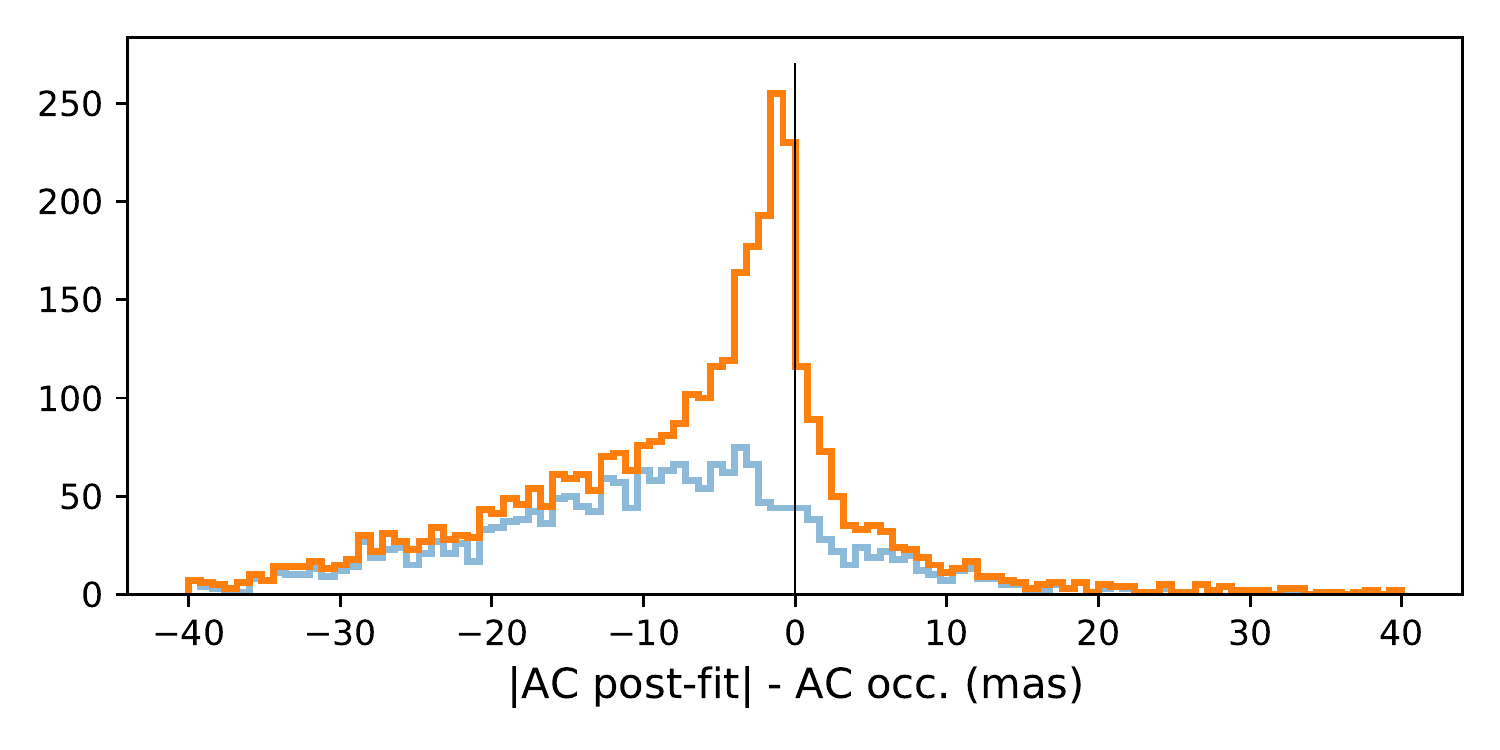}
\label{F:error_vs_residual}
\caption{Distribution of the difference between the post-fit residuals (absolute value) and the uncertainty attributed by the error model, for the orbits derived by occultations only. The distributions are for the directions AL (top) and AC (bottom). The histogram has 0.8 mas per bin.
In orange, the curve for all data, while residuals for single-chord occultations are shown in blue.}
\label{F:error_comparison}
\end{figure}

\begin{figure*}[ht]
\includegraphics[width=0.441\hsize]{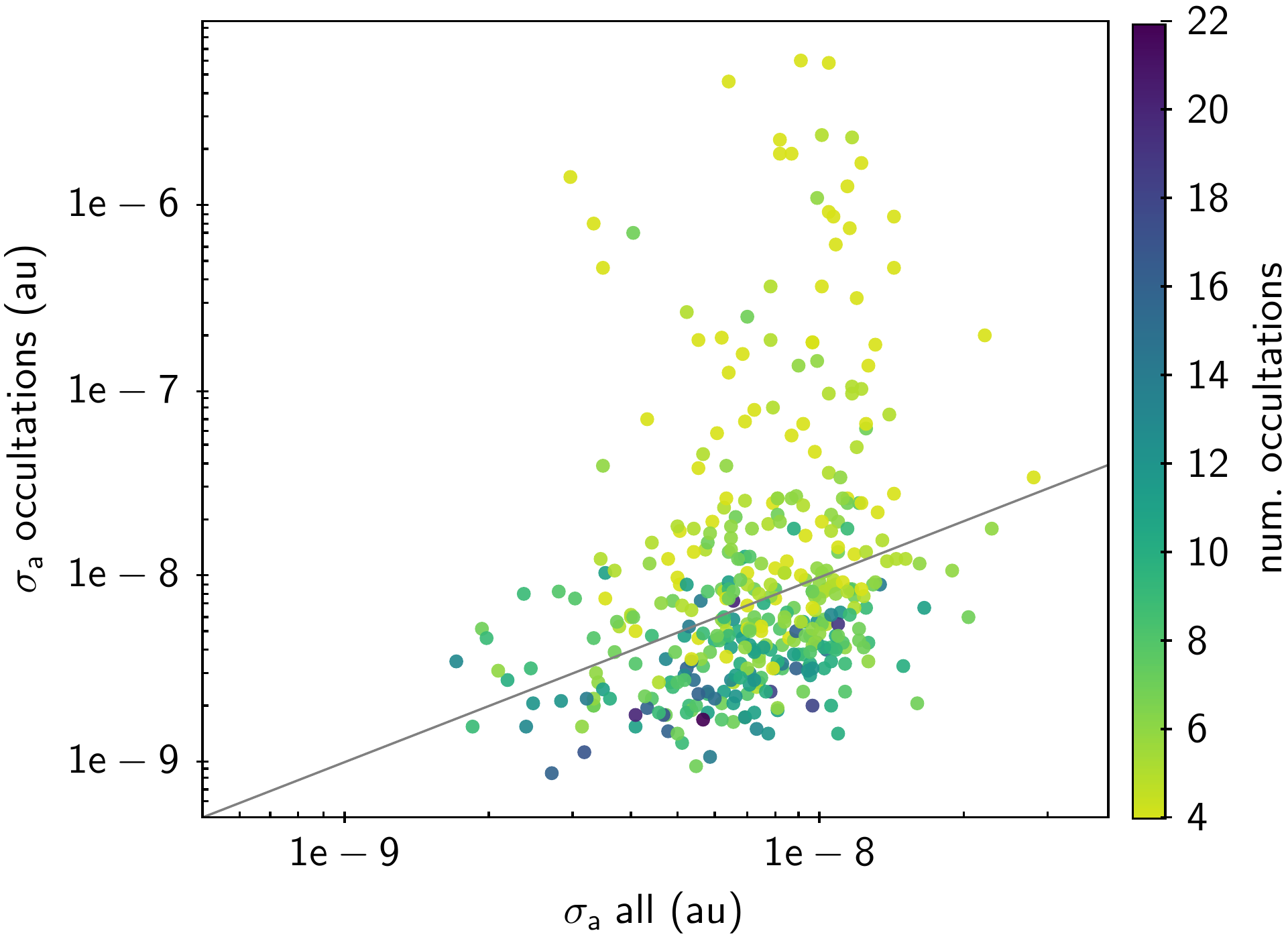}
\includegraphics[width=0.456\hsize]{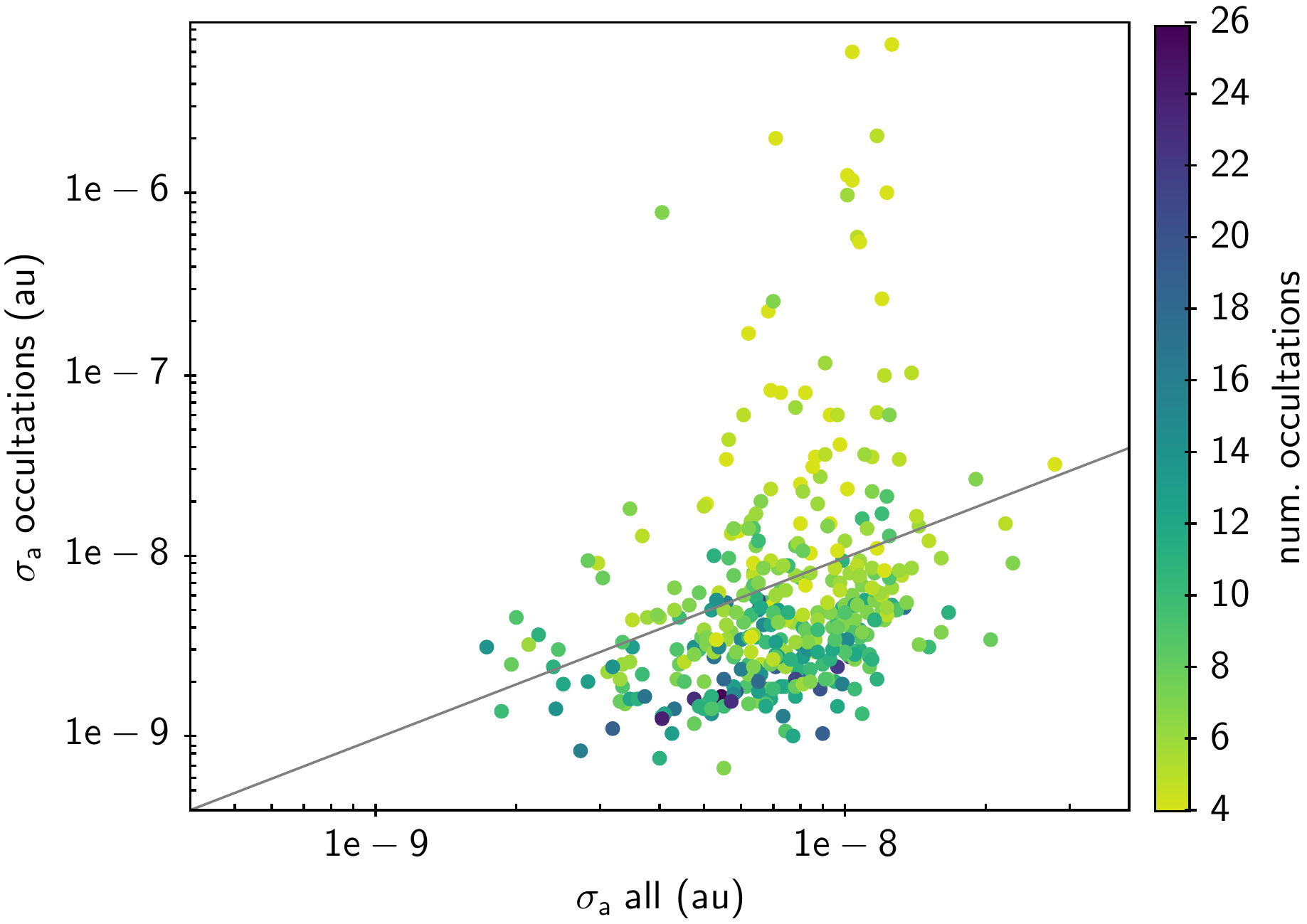}
\caption{Uncertainty on the semi-major axis for orbits obtained by occultations only, as a function of the same uncertainty but including all traditional observations. Each circle represents an asteroid, whose colour is related to the number of available occultations. The straight line is the bisector where the two uncertainties are equivalent. The right panel illustrates the situation at the time of DR2 release, to be compared with Fig.~6 in \citet{spoto_17}. The right panel corresponds to the same objects, but with the up-to-date situation and EDR3 stellar astrometry.}
\label{F:trad_vs_occ}
\end{figure*}

The comparison of the error model to the post-fit residuals (Fig.~\ref{F:error_comparison}) is helpful to evaluate the current uncertainty budget assigned to occultations. Given the possible additional sources of error, a perfect correspondence is not expected. The histograms generally show good agreement, with a peak at -2.3 and -0.8 mas, for the AL and AC directions, respectively, and a full width at half maximum of 4.0 and 5.5 mas. The shifted peak clearly indicates that even for multi-chord events the error model appears to be slightly conservative on average. One can also note that the histogram in AC is strongly skewed due to an overall overestimation of uncertainties for single-chord events. While these plots would suggest that the error model could probably be optimised, a conservative approach has the advantage of not putting excessive weight on single data points.




A major quality indicator for the fitted orbits is the uncertainty on the semi-major axis ($\sigma_a$). In Fig.~\ref{F:trad_vs_occ} we compare the uncertainties obtained from the fit of all available occultations to those of all other astrometric data, as in \citet{spoto_17}, Fig. 6, where occultations were reduced by \drone{}. An overall improvement of $\sim$1 order of magnitude with respect to the data reduction by \drone{} is clearly visible both when using \drtwo{} and \edrthr{}. As expected, the latter has the best performance, with a larger fraction of objects that have a better accuracy when considering occultations alone. We stress here that this primacy of occultation over the bulk of the other data is a proof of their quality and homogeneity. These properties derive not only from the excellent astrometric precision of each measurement, but also from the use of a single star catalogue (again, of the best quality) as a reference. This last issue is in striking contrast to the existing observation record that we used for comparison, including all other astrometric data, which are in general less accurate as they refer to a mix of astrometric catalogues and present systematic differences that cannot be completely suppressed.

\begin{figure}[ht]
\includegraphics[width=0.9\hsize]{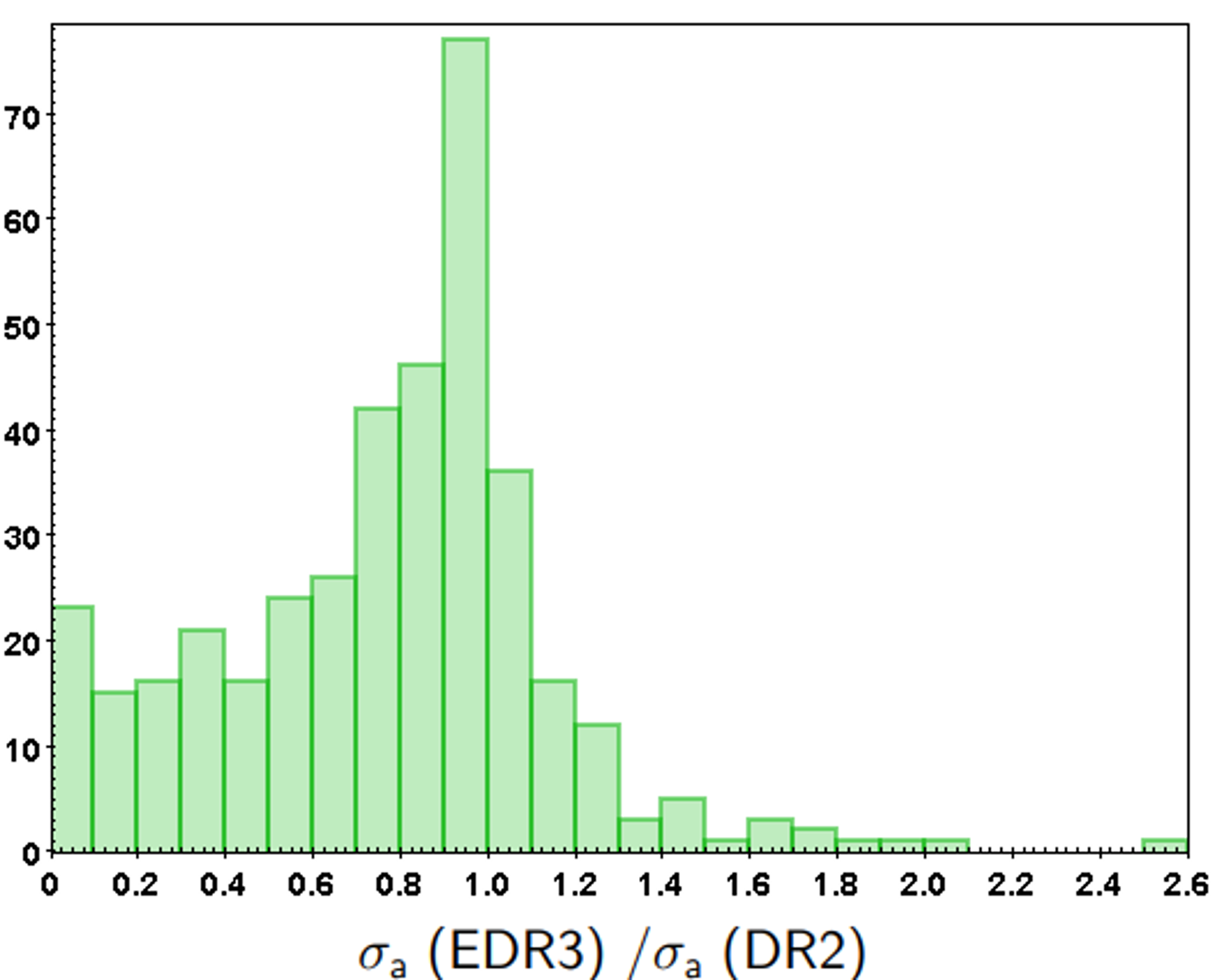}
\caption{Improvement factor of the semi-major axis uncertainty, between \drtwo{} and \edrthr{}, expressed by their ratio.}
\label{F:sigma_DR2_EDR3}
\end{figure}

Up to now, \drtwo{} and \edrthr{} have been used rather interchangeably in occultation prediction, without a clear, quantitative view of the improvement brought by the latter in the astrometric quality. It is possible to better appreciate the evolution of accuracy by Fig.~\ref{F:sigma_DR2_EDR3}, where the ratio of the semi-major axis uncertainty (same appearing on the vertical axis in Fig.~\ref{F:trad_vs_occ}) computed with the two data sets is presented. The distribution is centred around unity with a peak around 0.9, but the skewed distribution shows that a majority of semi-major axis uncertainties present an improvement, in some cases by a large factor. Despite the fact that statistics are small in the less populated bins, it appears that larger improvements occur preferentially for asteroids that have a small number of observed occultations. For them, an increase of accuracy on some of the few astrometric data points has a larger impact. We can thus say that the much refined treatment of both random and systematic errors in \edrthr{}, in particular for the bright sources (G$<$12) that dominate the sample of the occulted stars \citep{Lindegren20_EDR3}, produces a significant signature.  

We also present the residuals for three asteroids that have a good number of transits (Fig. \ref{F:singleobjects}), a low orbit uncertainty, and are representative of some typical situations. For all of them the residuals are plotted in the AL and AC directions. When examining the figures that report the absolute residuals in mas, as the size of the asteroid is different for each event, it is difficult to compare the quality of one observation with respect to the others. A given data point can be 'far' (compared to its error bars) from the centre, but if the asteroid's apparent size were much larger than this distance, it could still represent its astrometry well, and be compatible with the positive detection of the event.

To partially overcome this difficulty of interpretation, we also propose that the residuals are normalised to the object's apparent radius, computed at the epoch of each event. This is significant for occultations, as errors scale with size (especially in AC, but also AL for single-chord events in particular). In this case, the distance of the data points from the object and their error bars (which are also resized) can be compared to each other as they are at the same scale. By assuming a spherical shape, the circle represents the profile of the asteroid. It is easy to see that multi-chord events are well clustered towards the centre and never outside the circle. Of course, this is just a first-order approximation; shapes are not spherical in general and if an object is significantly elongated, with a different projected profile for each event, the interpretation is less straightforward. Eventually, errors on the stellar astrometry can also show up here under the form of larger residuals.

For (105) Artemis, the improvement with respect to \cite{spoto_17} is striking. All the measurements cluster very well within the circle representing the object size, with only one exception that remains compatible at the 1-sigma level.

For (25) Phocaea, the data are more scattered and the uncertainty larger, but overall they remain self-consistent. The data with the smallest errors clearly cluster very close to the object centre. Only one data point (at the top of the right panel) stands out as a clear outlier with respect to its very small error bars. Such a \textbf{situation} can be typical of a problematic star, which can adversely affect the quality of astrometry, even if the event is well observed. 

(176) Iduna is another case with a very good fit, and an astrometry more closely clustered in the AL direction, where the occultation timing is more sensitive. 

\begin{figure}[t]
\includegraphics[width=0.9\hsize]{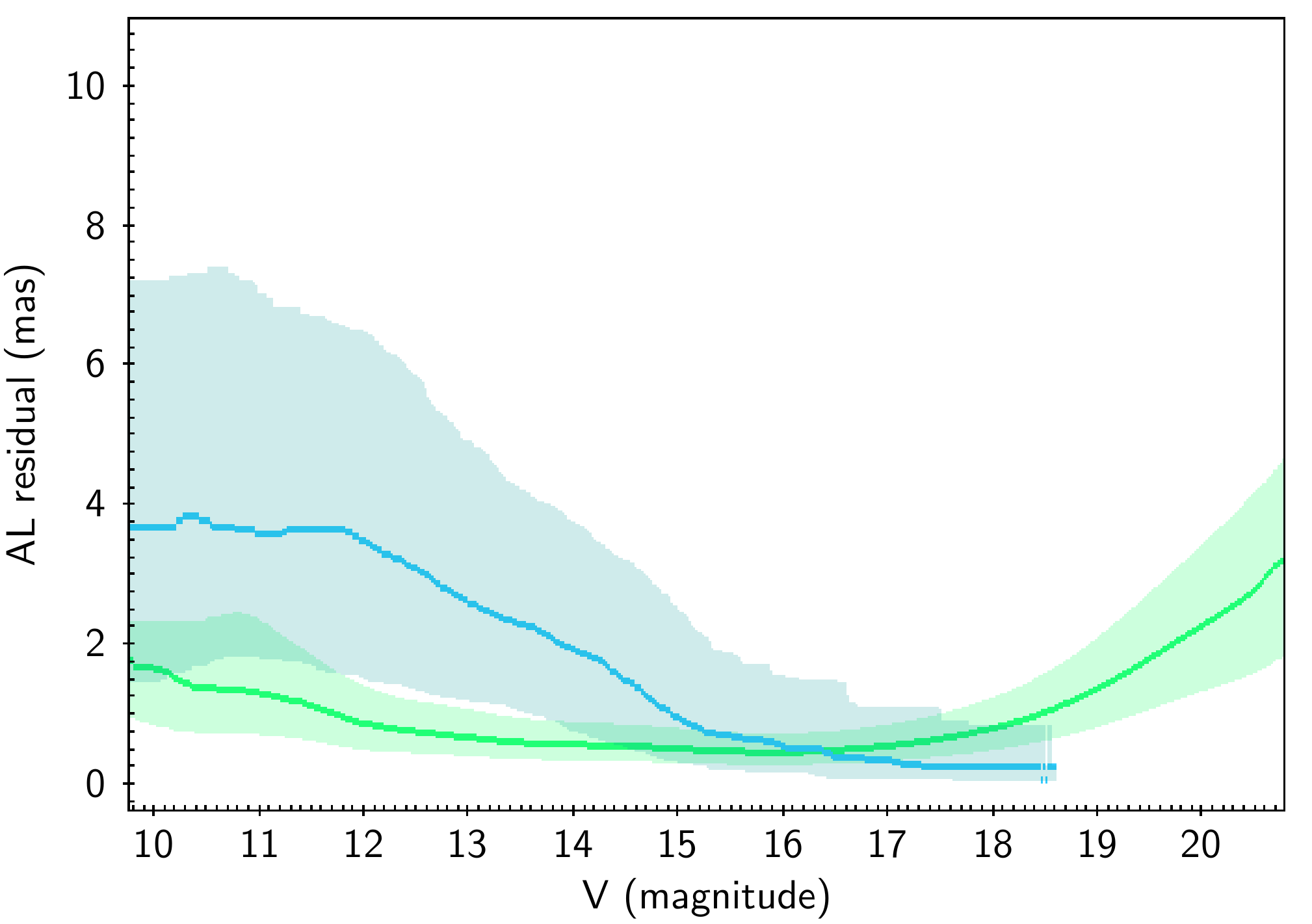}
\caption{Residuals with respect to an orbital fit to \drtwo{} asteroid astrometry (green), compared to a fit to occultations (blue). 510 asteroids for which the orbit have been computed are present in the \drtwo{} and are used to build the blue curve. The whole sample published in \drtwo{} is used for the green curve. Here, we consider residuals oriented along the two directions of better accuracy for both Gaia (along-scan direction) and occultations (AL). The curves represent the average, while the shaded areas encompass the 1-sigma level quantile. Statistics are not shown for V$<$10 due to the very reduced sample in that range.}
\label{F:occ_vs_DR2}
\end{figure}

\section{Discussion and perspectives}
\label{S:discuss} 

Our analysis clearly shows the potential of occultation astrometry and the advantage of referring its derivation to a unique catalogue reference of the best available quality, now provided by Gaia. The results obtained also clearly depend on the number of chords that are observed for each event, on the knowledge of the projected shape (often very approximate) and on the accuracy of the astrometry of the occulted star. These limitations are intrinsic to the observational approach but they are also the direct consequence of the increased resolution that occultations provide. We have also shown that while there is the possibility for some marginal improvements of the error model by \citet{Herald_2020}, as it is in general in good agreement with the residual. Optimisations, when possible, would probably require the application of more complex physical models of the object, with the danger of introducing other uncertainty sources. For the goal of orbital fitting, we can thus consider that the current error model is appropriate.

We now have all the elements required to provide a quantitative, statistical support to the idea that stellar occultations are capable to produce astrometry at the level of what Gaia can do for Solar System objects (i.e. mas-level epoch astrometry) and that occultations can provide a good continuation to Gaia in terms of astrometric precision even after its mission is over.

We illustrate this conclusion, without loss of generality, by adopting the point of view of traditional astrometry, and considering its performance as a function of the number of available photons. Data obtained from Gaia obey this law with an optimal performance around G$\sim$12-13, and a degradation for decreasing brightness. 

In the case of epoch astrometry of asteroids in \drtwo{}, the optimal performance is shifted towards G$\sim$15, showing an additional difficulty of Gaia with regard to dealing with bright and marginally resolved sources \citep{Spoto_2018}. Of course, the decrease in accuracy at fainter magnitudes is also clear for asteroids. This magnitude dependence appears with a completely opposite signature for occultations as, on average, fainter asteroids are smaller and thus result in errors due to size or shape effects that proportionally decrease in their amplitude. 

The difference in behaviour between \drtwo{} asteroid astrometry and occultations is illustrated in Fig.~\ref{F:occ_vs_DR2}, where we show the residuals with respect to the orbital fitting, for both data sets, after selecting only the asteroids in common. For each occultation or Gaia epoch, the computed apparent V magnitude of the object is used to build the plot. One should note that we limited the comparison to AL for both occultations and Gaia. In fact, Gaia astrometry is essentially mono-dimensional along the instantaneous direction of its scanning motion, and brings nearly meaningless constraints to the object positions in the AC direction.

The first clear feature of the plot is the generally better performance of \drtwo{}, but with occultations that have (on average) residuals only two to four times larger. 
However, occultations also have a much larger dispersion at the bright end, where errors associated with the asteroid size and shape for large asteroids, sampled by a small number of chords, show up. Smaller flux drops associated with occultations by large asteroids can also contribute to increase the timing uncertainty; however, a careful verification showed that it is a minor effect. Only 3.2$\%$ of the occultations involving a star fainter than the asteroid (representing 17$\%$ of the whole occultation data set) have a timing error larger than errors associated to size and shape. Such errors of course diminish with smaller asteroids so both the average and the dispersion decrease, as expected, until they surpass Gaia in performance around V$\sim$16.3 for the AL direction.

Although the idea of 'doing better than Gaia' can appear surprising in itself, we stress here that there is no contradiction when the quantities that compared are correctly interpreted. In fact, on one hand we have the \drtwo{} astrometry of asteroids for each epoch. On the other hand, we have the occultation astrometry based on \edrthr{} stellar astrometry. The latter is obtained from the combination of all the epochs of observation of each star, and, as a consequence, it is intrinsically better than for asteroids. It is thus natural that occultation astrometry has an error floor that is set by the stellar astrometry of Gaia, not by the performance of single-epoch asteroid astrometry. The decreasing errors of occultations appear to approach that floor towards the faintest end of their available range. By observing more occultations of small asteroids in the future, the curve should start to increase again towards fainter magnitudes V$>$17. This transition, assuming an average albedo of 0.15 and a main belt asteroid (semi-major axis between 2.2 and 3.2 au), corresponds to a size approximately in the 5-8 km range.

Finally, we briefly comment the issue of combining occultation data to the large amount of astrometric data available. In the end, this is required since accurate occultation data for each object are limited to a small number of data points in general, and in some situations they are not sufficient to reach an adequate performance in terms of orbit accuracy. Also, if any systematic effect is hidden in the few occultation data available, a high orbit accuracy cannot exclude a biased orbital solution. In the right panel of Fig.~\ref{F:post2000}, we show residuals obtained from the combination of traditional astrometry and stellar occultations. Of course, the large majority of traditional astrometric data dominates the distribution. Only a careful investigation, taking into account the change in orbital quality due to occultations, as a function of the number and quality of occultation events, the volume of pre-existing astrometry, and so on, can reveal the degree of improvement brought by a very small fraction of ultra-accurate astrometric data.

In this respect, we find again the classical problem of orbit optimisation, starting from a sample of heterogeneous observations, with different error distributions, affected by a variety of different biases. While it is outside the scope of this article to propose approaches to improve the result of this comprehensive fit, we stress here the importance of this further step, which is required to correctly couple the best asteroid astrometry available, to the large existing sample of observations. The recent series of successful observations of very difficult predictions, spanning from the far TNO (486 958) Arrokoth to the small NEAs (3 200) Phaethon and (99 942) Apophis, and to specific TNO/Centaurs\footnote{Often in the frame of large collaborations such as the Lucky Star project \url{https://lesia.obspm.fr/lucky-star/index.php}}, have been granted by the success of such optimisation procedures, often requiring a careful data selection and error modelling \citep[see e.g.][]{Porter2018, arrokoth, Rommmel}.

In conclusion, despite the fact that the full exploitation of the potential of occultation astrometry on specific targets remains a challenging task, we have shown the impressive statistical properties of occultation astrometry. This data set is now growing rapidly and we can expect a further acceleration of successful predictions, towards more numerous events for smaller asteroids, in the coming years, as a consequence of the forthcoming data releases. As discussed, occultation astrometry is not far in performance from what Gaia can obtain, and it will thus represent an impressive extension in time of ultra-accurate astrometry on all categories of minor bodies in the Solar System.

\begin{figure*}[h!]
\begin{center}
\includegraphics[width=0.38\hsize]{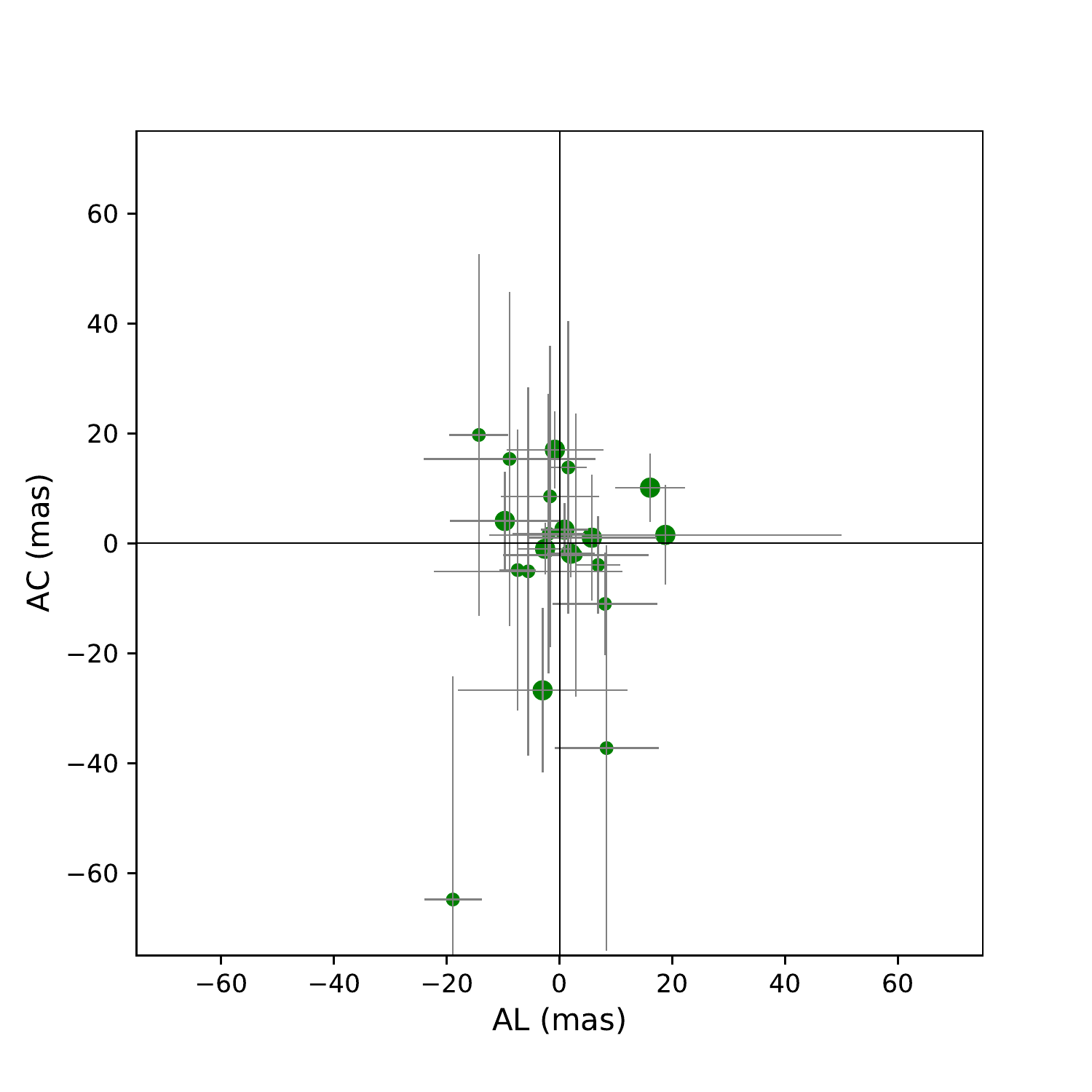}
\hspace{0em}
\includegraphics[width=0.38\hsize]{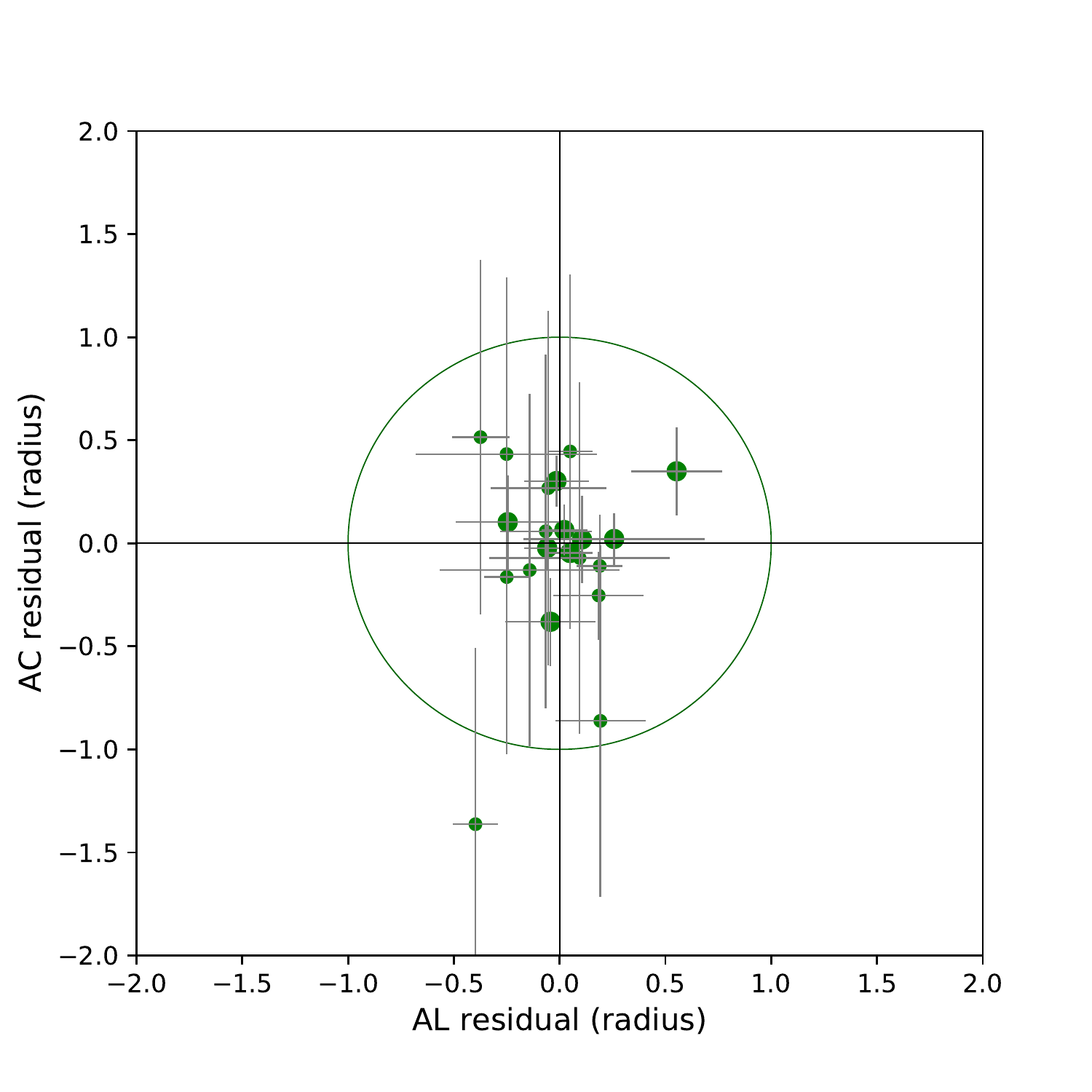}
\includegraphics[width=0.38\hsize]{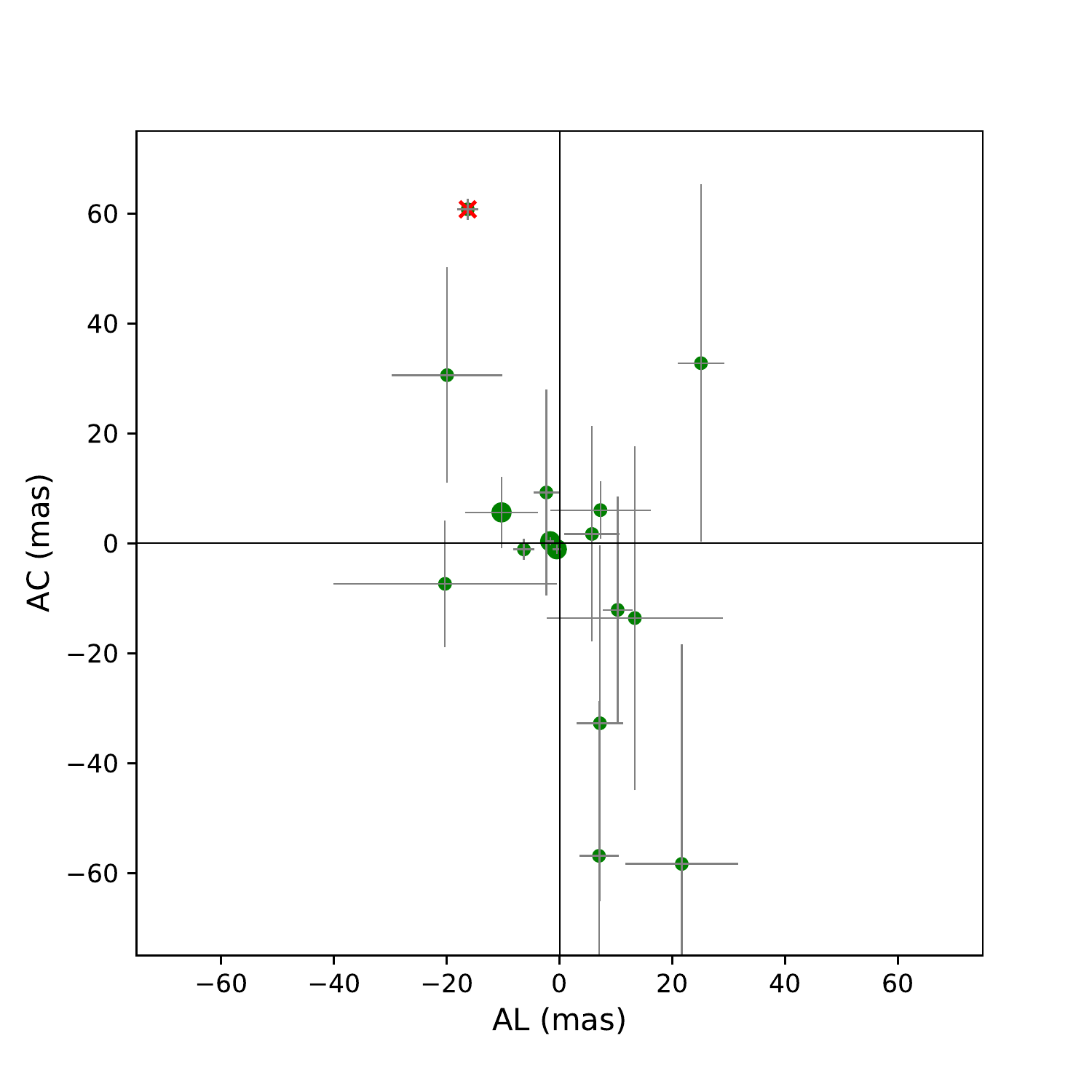}
\hspace{0em}
\includegraphics[width=0.38\hsize]{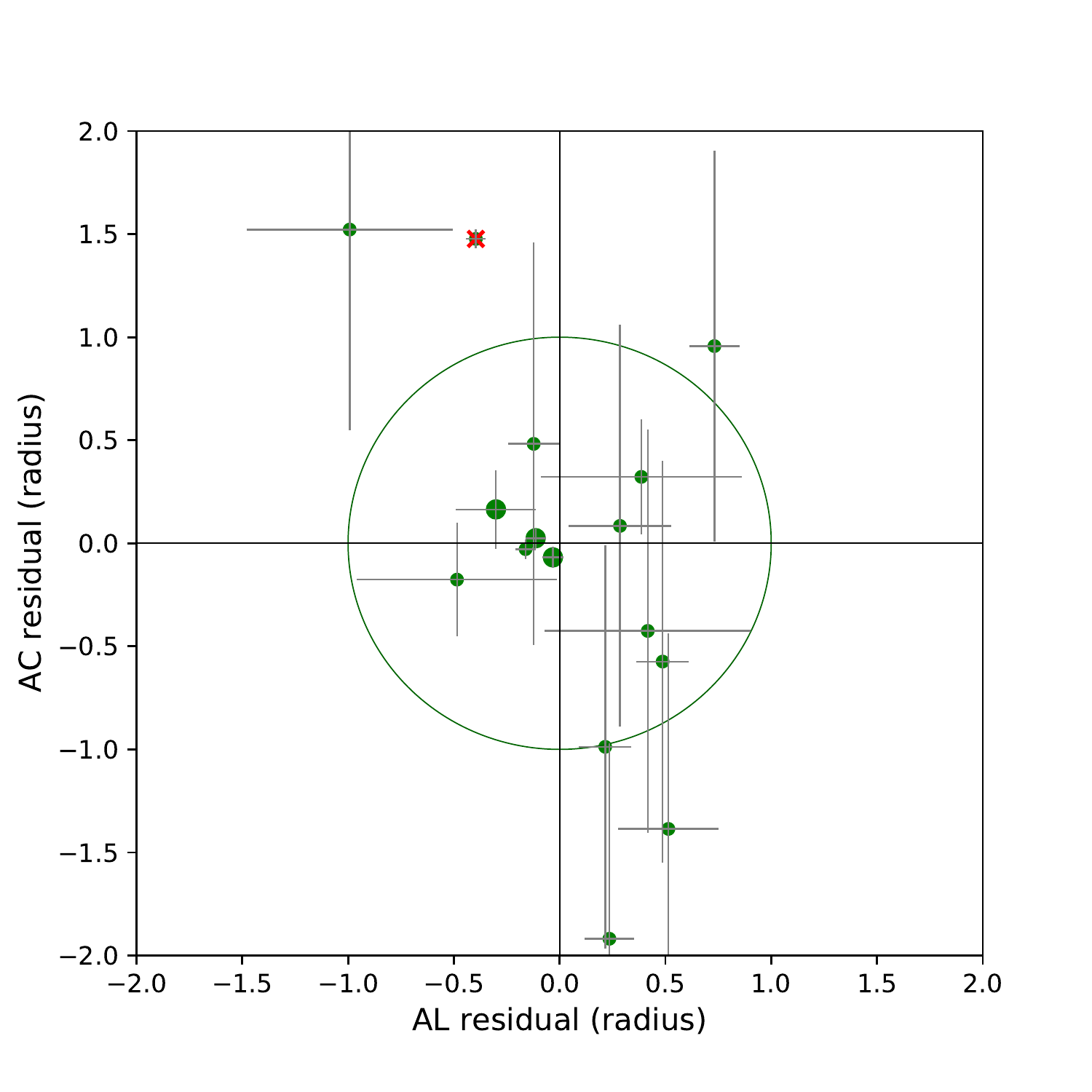}
\includegraphics[width=0.38\hsize]{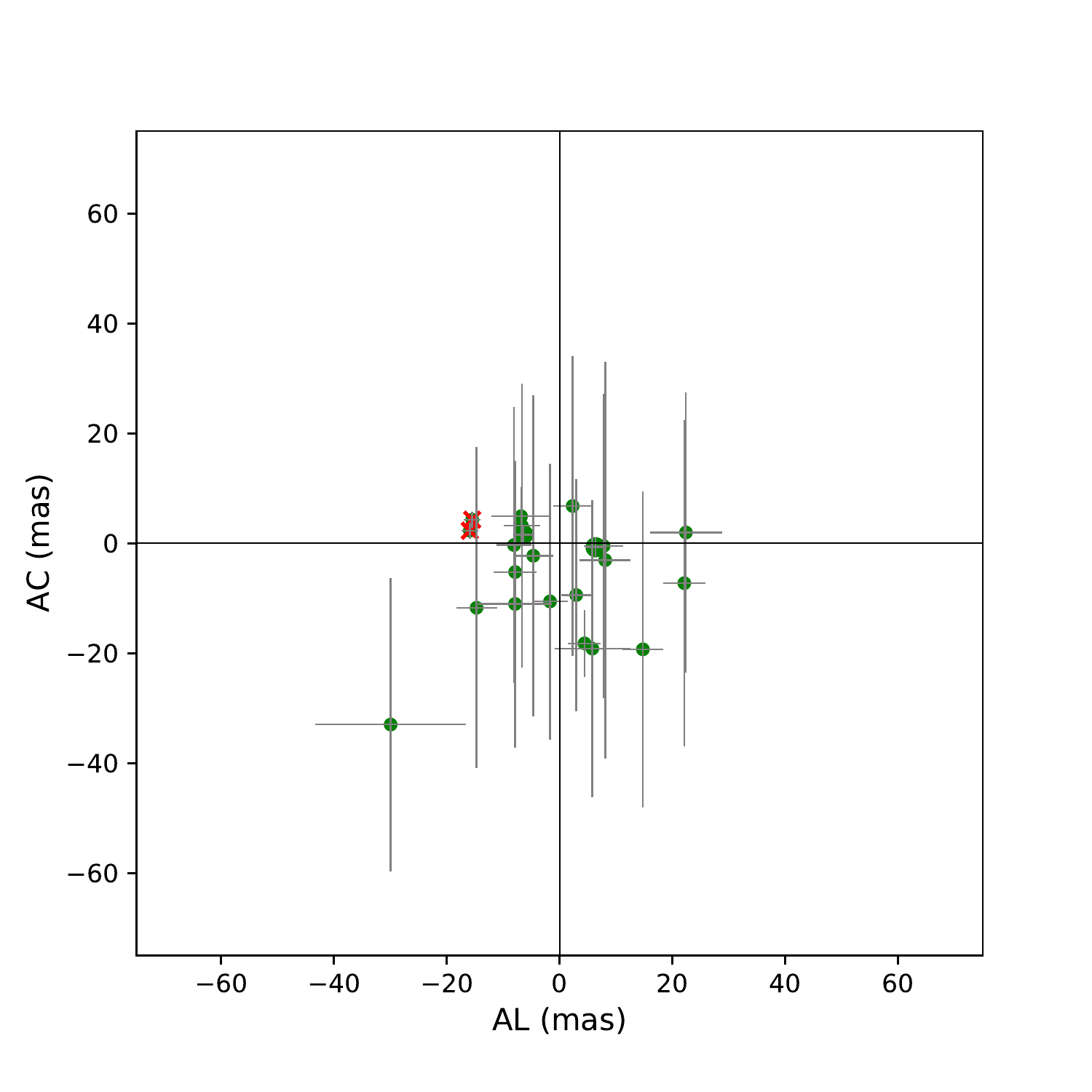}
\hspace{0em}
\includegraphics[width=0.38\hsize]{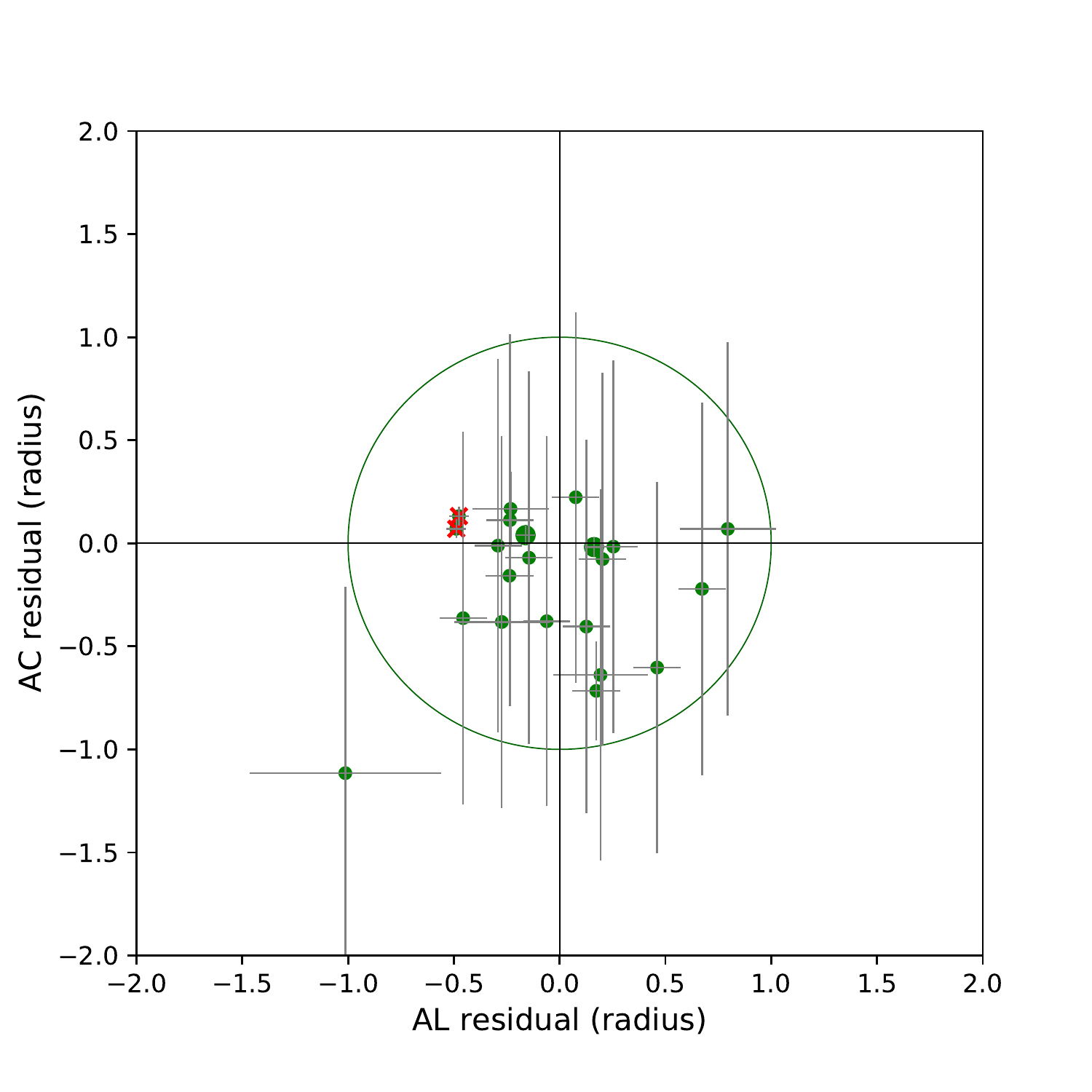}
\end{center}
\caption{Post-fit residuals on the (AL, AC) plane for (105) Artemis (top row), (25) Phocaea (middle), and (176) Iduna (bottom row). The error bars represent the 1-sigma level of the uncertainties assigned by the error model. In the right panel, they have been normalised to the apparent average radius of the asteroid at the epoch corresponding to each observation. The corresponding circle is shown. The red crosses represent observations that are rejected by the orbital fitting process. Larger symbols are used for events with $>$2 chords. The plots for (105) are compared to Fig.~7 in \citet{spoto_17}.}
\label{F:singleobjects}
\end{figure*}

\section*{Acknowledgements}
    
This research has been supported by the Pessoa program for science cooperation between Portugal and France, the {\sl Programme Nationale de Planetologie} in France, the BQR programs of Laboratoire Lagrange and Observatoire de la C\^ote d'Azur. 

PM acknowledges support from the Portuguese Fundação Para a Ciência e a
Tecnologia project P-TUGA Ref. PTDC/FIS-AST/29942/2017 through national
funds and by FEDER through COMPETE 2020 (Ref. POCI-01-0145 FEDER-007672).

We made use of data from the European Space Agency (ESA) mission {\it Gaia} (\url{https://www.cosmos.esa.int/gaia}), processed by the {\it Gaia} Data Processing and Analysis Consortium (DPAC, \url{https://www.cosmos.esa.int/web/gaia/dpac/consortium}). Funding for the DPAC
has been provided by national institutions, in particular the institutions participating in the {\it Gaia} Multilateral Agreement.

This research made use of Astropy,\footnote{\url{http://www.astropy.org}} \citep{astropy:2013, astropy:2018}; matplotlib \citep{Hunter2007}; scipy \citep{scipy2019}.

We acknowledge the anonymous referee whose contribution was essential to improve our article. 

    \bibliographystyle{aa} 

    \bibliography{mybib}
\end{document}